\documentclass[superscriptaddress,floatfix,twocolumn,showpacs]{revtex4}
\usepackage{graphicx,bm,epstopdf}
\usepackage{amsmath}
\usepackage{fancybox}

\newcommand{\ie}{\emph{i.e.}}
\newcommand{\eg}{\emph{e.g.}}

\begin{document}

\title{Modeling and verifying a broad array of network properties}
\author{Vladimir Filkov}
\email{filkov@cs.ucdavis.edu}
\affiliation{Department of Computer Science, University of California, Davis, CA 95616}
\author{Zachary M. Saul}
\affiliation{Department of Computer Science, University of California, Davis, CA 95616}
\author{Soumen Roy}
 \affiliation{Center for Computational Science and Engineering, University of California, Davis, CA 95616}
 \affiliation{Department of Mechanical and Aeronautical Engineering, University of California, Davis, CA 95616}
\author{Raissa M. D'Souza}
\affiliation{Center for Computational Science and Engineering, University of California, Davis, CA 95616}
\affiliation{Department of Mechanical and Aeronautical Engineering, University of California, Davis, CA 95616}
\author{Premkumar T. Devanbu}
\affiliation{Department of Computer Science, University of California, Davis, CA 95616}
\def\sa{self-averaging }
\def\begeq{\begin{equation}}
\def\endeq{\end{equation}}
\def\bege{\begin{eqnarray}}
\def\ende{\end{eqnarray}}

\begin{abstract}
Motivated by widely observed examples in nature, society and software,  where
groups of related nodes arrive together and attach to existing
networks, we consider network growth via sequential attachment of linked node groups,
or graphlets.  
We analyze the simplest case, attachment of the three node \hbox{$\bigvee$-graphlet},
where, with probability $\alpha$, we attach a peripheral node of the graphlet, and
with probability $(1-\alpha)$, we attach the central node.  Our analytical
results and simulations show that tuning $\alpha$ produces a wide range in
degree distribution and degree assortativity, achieving assortativity values
that capture a diverse set of many real-world systems.  We introduce a
fifteen-dimensional attribute vector derived from seven well-known network
properties, which enables comprehensive comparison between any two networks.  
Principal Component Analysis of this attribute vector space shows 
a significantly larger coverage potential of real-world network 
properties by a simple extension of the above model when compared 
against a classic model of network growth.
\end{abstract}

\maketitle
\begin{figure*}
\hfill\hfill
\begin{minipage}{0.3\textwidth}
\includegraphics[width=\columnwidth]{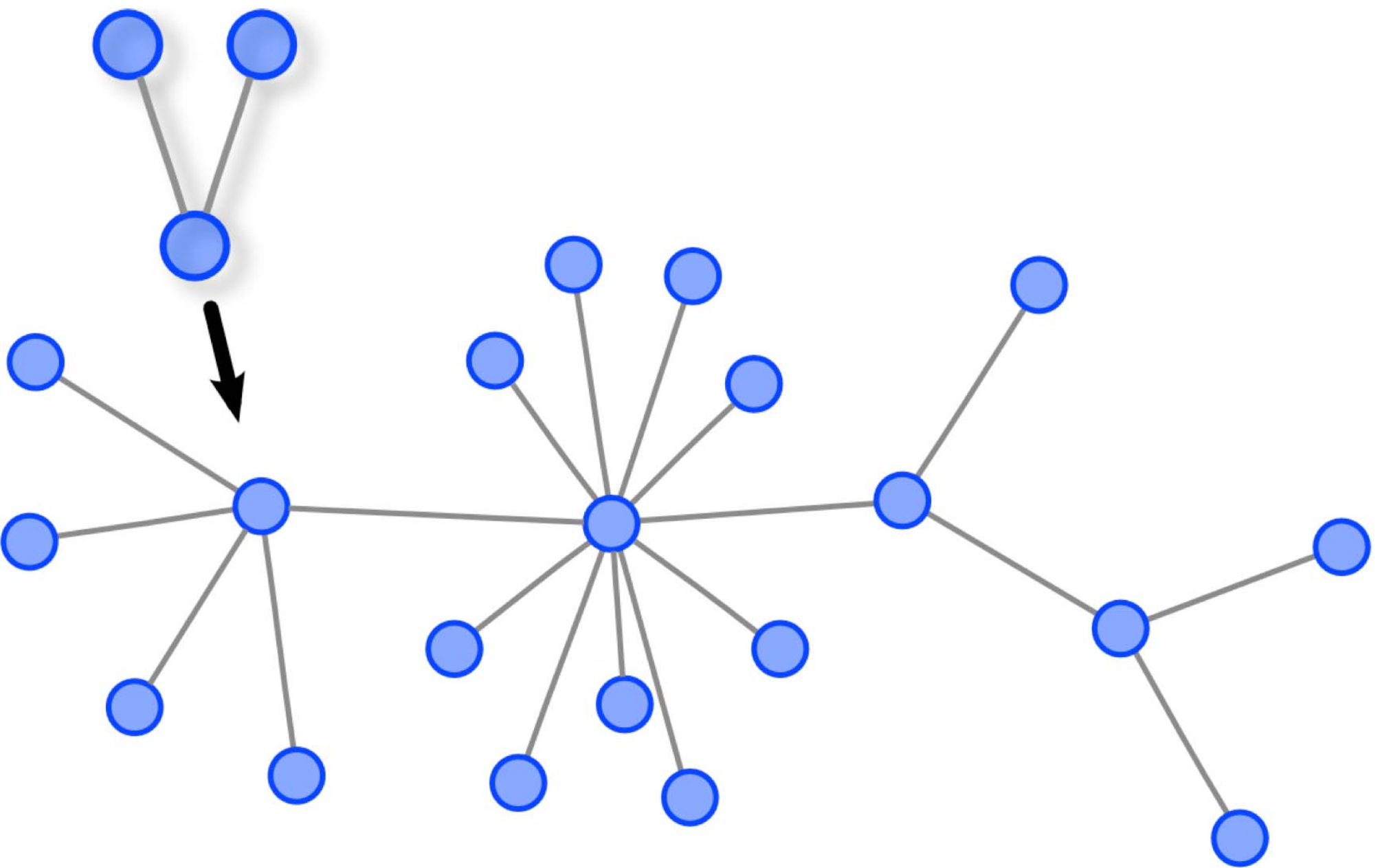}
\end{minipage}
\hfill
\begin{minipage}{0.5\textwidth}
\includegraphics[width=\columnwidth]{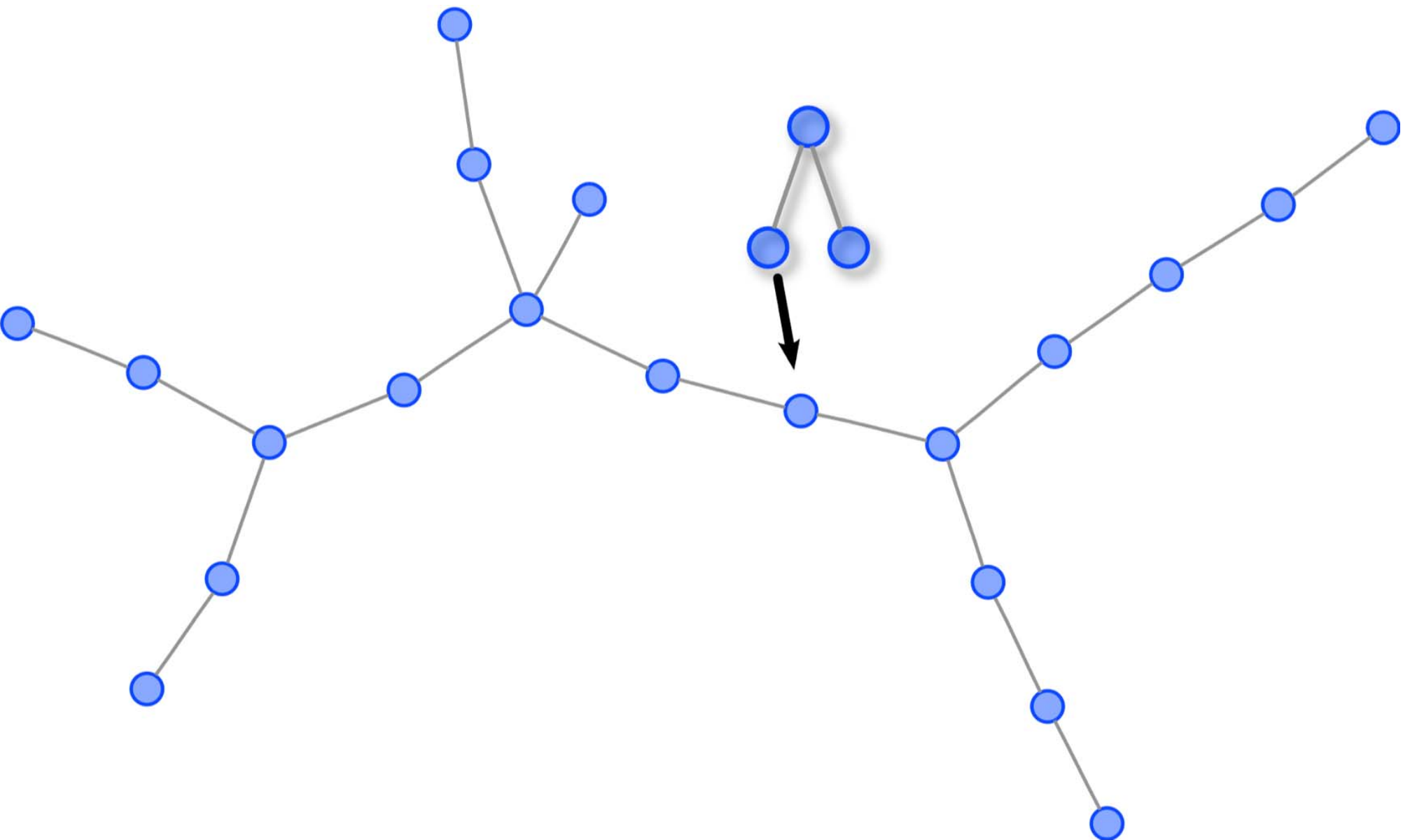}
\end{minipage}
%
\caption{
\label{figure:growth}
The Growth Process.  The $\bigvee$-graphlet arrives and merges into the
existing network at either its midpoint (with probability $1-\alpha$) or its periphery
(with probability $\alpha$).  Here we show the process, after
the arrival of 10 graphlets for $\alpha=0$ (left) and  $\alpha=1$ (right).
Already the creation/surpression of hubs is evident as well as the more
homogeneous nature of the degree distribution for $\alpha=1$. 
Networks grown with $0 < \alpha < 1$ show behaviors intermediate between
these two.
}
\end{figure*}

The ubiquity and importance of network structures has recently become apparent,
leading to an increased focus on network growth mechanisms~\cite{albert:02smo}. 
 Existing models of network
growth primarily consider the arrival of single nodes at each time step;
however, there are numerous examples in natural and artificial systems where
networks grow not just by the addition of single nodes but by the addition of
groups of already related nodes.  For example, in biology, in developmental transcriptional gene regulation,
whole pathways can be added or eliminated by a mutation in a master
regulator~\cite{davidson06}; and in the evolution of biological networks, gene
duplication can add subnetworks to the network~\cite{weitz2007}.  
Growth of computer software networks (composed of interacting functions or classes) is often due
to adding small groups of related elements simultaneously.  For example,
1)~functions to allocate, use, and free a resource (such as a file) are usually
added together and 2)~in object-oriented languages, good design principles call
for classes to be added in small groups called \emph{design
patterns}~\cite{gamma1995dpe}.  Further, in social networks within cities,
families arrive as units, and growth can be described via aggregation of small
pre-existing modules.  Similarly, in corporate enterprises, the practice of
``lift-outs", employing pre-existing functional teams of people (rather than
building up a team from individual hires), is on the rise~\cite{liftouts:HBR06}.  
This insight suggests that a new class of network growth models incorporating group arrival could lead to more realistic models.
Moreover, most existing work in modeling network growth focuses on matching a single or 
few attributes of 
empirical networks, in particular degree distribution, clustering coefficient, etc. 
But networks can differ in many ways while being similar in others, e.g. some with the same degree 
distribution have different levels of assortative mixing. Thus, a more comprehensive comparison, simultaneously across many important attributes, is desirable.

Hence, the purpose of this letter is two-fold.  Our  first intent is to 
propose modeling network growth by sequential aggregation of groups of nodes, represented by
small, connected graphs or {\em graphlets} attaching preferentially in the network, rather than 
by preferential attachment of single nodes.  Thus, we introduce the {\em graphlet arrival model} and show that in spite of its added complexity important analytical results can be obtained.
 The model based on iteratively adding the three-node $\bigvee$-graphlet 
yields networks with degree distributions (the distribution of the
probability of observing a node of degree $k$) that follow an asymptotic power
law, \ie, $p_k \sim k^{-\gamma}$, where, $\gamma$ is a parameter ranging from $3 \le \gamma \le 5$, 
in agreement with those found in a number of highly-cited studies of  
real-world systems where graphlets could play a crucial role
~\cite{semantic:steyvers}.
We also analytically derive  the {\it degree assortativity}, $\rho$, 
a measure of the tendency of nodes to link to nodes of like degree, which has
the power to discriminate between empirical networks from various fields, 
 even if they have similar degree distributions~\cite{newman:02ami, newman:03wsn}.  
As noted recently~\cite{newman:02ami}, an interesting open problem is to come up with a single growth model 
which could generate networks of both positive assortativity, like social networks, and negative assortativity, like technological and biological networks.
 We find that our model yields tunable assortativity, with respect to a parameter, $\alpha$, which determines the graphlet attachment point probability, as explained below.  
Our numerical results for networks up to $\simeq10^7$ nodes in size (which covers most real-world networks) show assortative behavior, ($\rho > 0)$, for lower $\alpha$ and dissortative behavior, ($\rho<0$), for higher values of $\alpha$.
Our  analytical calculations show that $\rho \ge 0$ for infinite size networks.

The second intent of this letter is to introduce techniques for
comprehensively comparing networks across a suite of network properties {\it simultaneously}, 
allowing for a much  more in-depth evaluation of network models 
than is possible using the commonly existing practice of comparing primarily degree distribution.
To that end, we compare the ability of our model networks to match the variability of $113$ real networks under $15$ attributes, and demonstrate how  data mining methods like clustering and 
statistical dimension reduction (via Principal Component Analysis) can be utilized to assess that match. 
A simple extension of our model yields remarkably large coverage of the attribute space spanned by the $113$ real networks, and a significant match of the ranges of real networks over all attributes.

To fully model with the graphlet arrival paradigm, one must
decide on which graphlet(s) to use, with which of their nodes to attach, and
where in the network to attach them. 
Common undirected graphlets include the dyad (edge), the two triads ($3$ nodes) and
the six tetrads ($4$ nodes). 
 To properly analyze their arrival and attachment into the network
one must classify the graphlets' nodes into equivalence classes
based on symmetry. 
Our model, illustrated in Fig.~\ref{figure:growth}, considers the simplest non-trivial case:  series of arriving triads consisting of a single node of degree two and two identical nodes of degree one, which we call the $\bigvee$-graphlet. 
This graphlet's asymmetry provides a choice of two topologically different attachment points (the two nodes of degree one are equivalent but different than the single node of degree two), unlike the edge and triangle graphlets which allow only one. 
The graphlets attach to the network by merging one of their vertices into an existing node
selected with probability proportional to the node's degree, i.e., via preferential attachment. 
The model chooses the
degree-one merge point with probability $\alpha$ and the degree-two merge point with probability $(1-\alpha)$.

\begin{figure}
\boxput*(-0.15,0.37) { \includegraphics[width=0.57\columnwidth]
{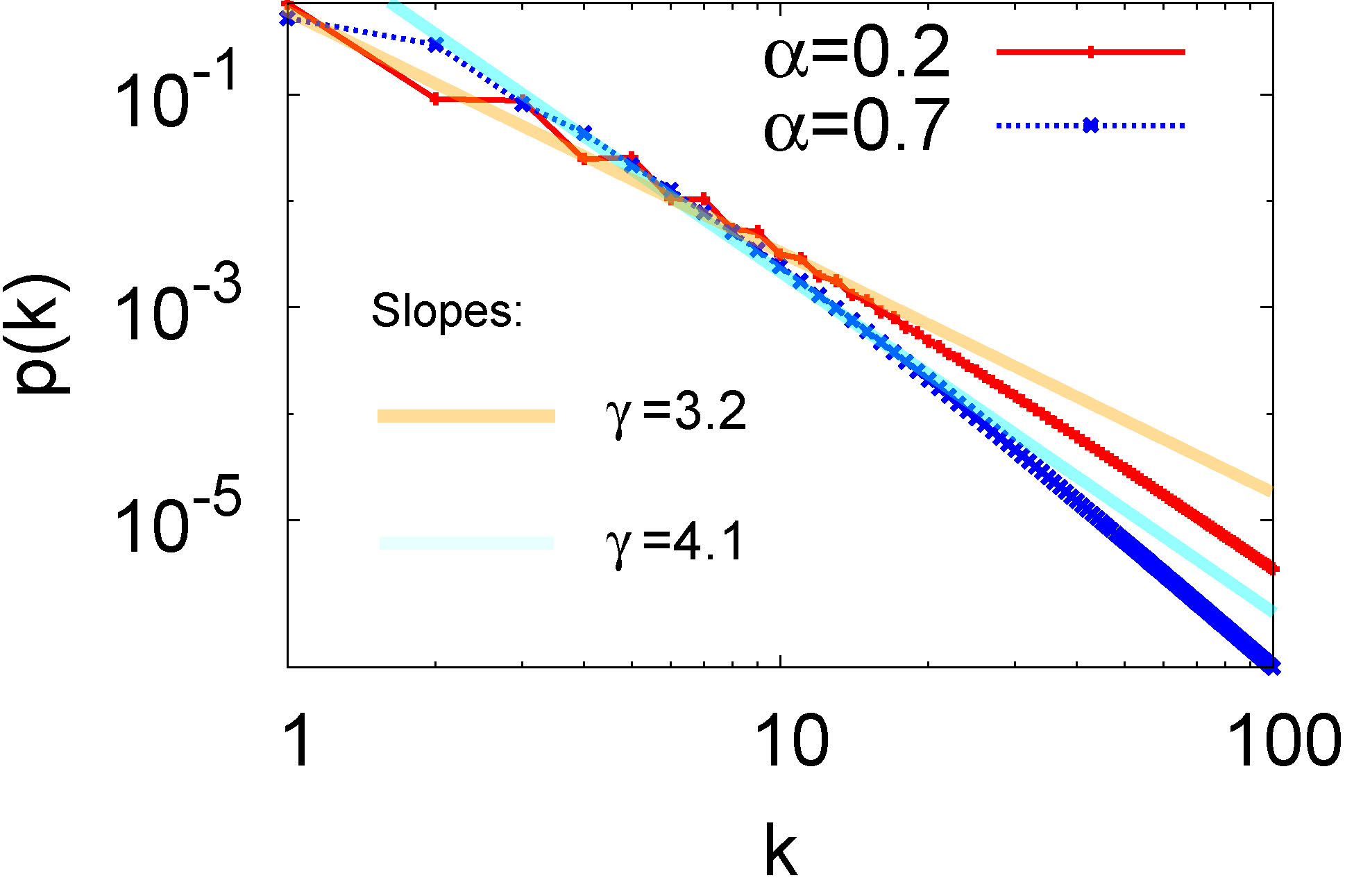} } 
{ \includegraphics[width=0.95\columnwidth]{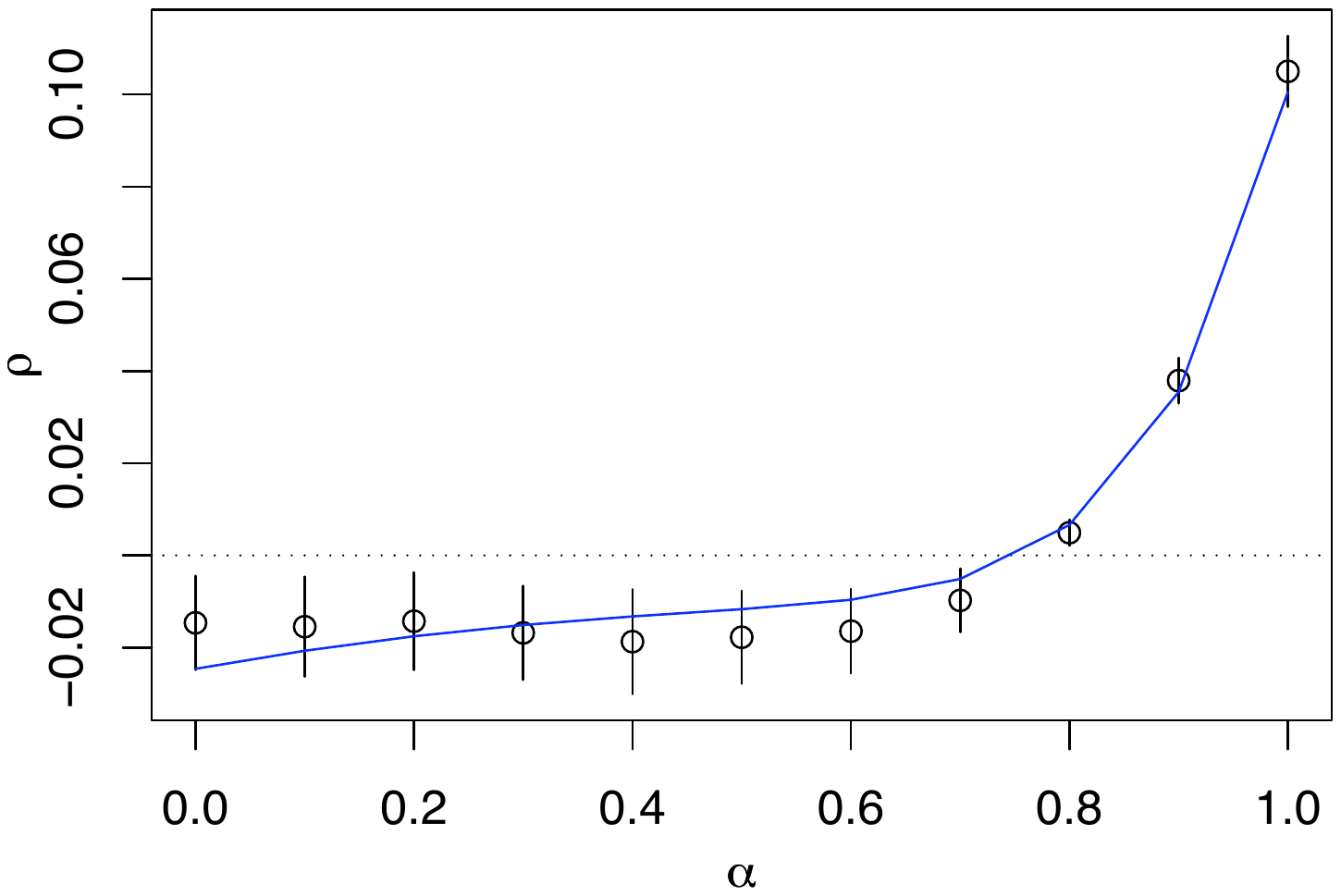} }
\caption{
\label{figure:dd-assort}
(Color online) Points show the mean $\rho$ over 100
simulations of $10^6$ node networks. Bars represent values within two standard
deviations. Solid line is the theoretical prediction for networks with maximum
degree $2500$. Inset: Distribution of average degree for $\alpha = 0.2$ and $\alpha=0.7$ over an 
ensemble of 5000 realizations,
 together with best fitting lines with slopes equal to the, respective, 
analytical $\gamma$'s of $3.2$ and $4.1$.}
\end{figure}

First, we derive the asymptotic degree distribution, $p_k$, 
for the $\bigvee$-graphlet arrival model via a master equation approach.
Starting with a single edge at time $t=0$, the number of nodes
at time $t$ is $N(t) = 2t + 2\approx 2t$, for large $t$.  Let $d_i(t)$ denote
the degree of vertex $i$ at time $t$. Then, the
probability that incoming graphlet $j$ merges with node $i$ is
%
$p_{j \rightarrow i} = \frac{d_i(t)}{\sum_i d_i(t)} = \frac{d_i(t)}{2N(t)} = \frac{d_i(t)}{4t}, $
%
where $\sum_i d_i(t) = 2N(t)$ as  
there is one edge for each node in the graph.  
Let $N_k(t)$ be the number of nodes with degree $k$ at 
time $t$.  Due to the asymmetry of the $\bigvee$, we get  
separate equations of $N_{k}(t)$ for $k\ge 3$, $k=2$ and $k=1$.
Making the natural assumption that $p_k(t) = N_k(t)/N(t)$ and assuming steady-state $(p_k(t) \rightarrow p_k)$ leads to $N_k(t) = 2t p_k$. 
From this and the $N_k(t)$ equations, which may be detailed elsewhere ~\cite{unpublished}, we get :
\begin{equation*}
\label{refeqn}
p_{k\ge3}  =  \alpha [(k-1)/(k+4) ]p_{k-1} + (1-\alpha) [(k-2)/(k+4)] p_{k-2},
\end{equation*}
with  $p_2  =  \frac{\alpha}{15}(7-\alpha)$ and $p_1 = \frac{2}{5}(2-\alpha)$.
Since $p_{k\ge3}$ 
depends on both $p_{k-1}$ and $p_{k-2}$ non-trivially,
we cannot solve it analytically. However for large $k$, a simple linear approximation results in $\gamma = (6-\alpha)/(2-\alpha)$. 
The results of numerical solutions are shown in the inset to
Fig.~\ref{figure:dd-assort}. 

The degree assortativity, $\rho$, is defined as the Pearson 
correlation coefficient between
the degrees of all pairs of connected vertices in the network~\cite{newman:02ami}.
Here, using a rate equation approach~\cite{chkns,krapivsky2001}, we directly calculate $\rho$ from $e_{kl}$, the probability distribution that an edge in an undirected graph is incident to vertices of degree $k$ and $l$, and $p_k$, the degree distribution 
~\cite{unpublished}.
Let $E_{kl}(t)$ denote the number of edges with a vertex of degree $k$ at one end and a vertex of degree $l$ at the other at time $t$.
We note that $\sum_{k \ge l}{E_{kl}(t)}=2 t + 2 \approx 2 t$, for large $t$, 
which implies $E_{kl}= 2 t e_{kl}$, for steady state. 
To derive a rate equation for $E_{kl}(t)$ we account for the processes that
change it when a new $\bigvee$ arrives. The processes
that increase $E_{kl}$ are when: with probability $(1-\alpha)$, a $\bigvee$ 
merges its midpoint to a vertex of degree $k-2$, which is already attached to a
vertex of degree $l$ (and the same argument with $k$ and $l$ reversed); with
probability $\alpha$, a $\bigvee$ merges one of its endpoints to a vertex
of degree $k-1$ (respectively $l-1$), which is already attached to a vertex of
degree $l$ (respectively $k$); in the special case when $k=1$, with probability
$(1-\alpha)$, a $\bigvee$ merges its midpoint to a vertex of degree $l-2$,
producing two new edges, each incident to vertices of degree $l$ and $1$; 
in the special case when $k=2$, with probability $\alpha$, a $\bigvee$
merges one of its endpoints to a vertex of degree $l-1$, producing one new edge 
incident to vertices of degree $l$ and $2$.  
The processes that decrease $E_{kl}$ are when: with probability
$(1-\alpha)$ a new $\bigvee$ merges its midpoint to a vertex of degree $k$
(respectively $l$), which is already attached to a vertex of degree $l$
(respectively $k$); with probability $\alpha$ a new $\bigvee$ merges one of its
endpoints to a vertex of degree $k$ (respectively $l$), which is already
attached to a vertex of degree $l$ (respectively $k$).  
From these cases, 
and incorporating preferential attachment 
(by multiplying the number of edges gained or lost by
$m/{4t}$, where $m$ is the degree of the node to which the new
$\bigvee$ is attached), we derive a rate equation for $E_{kl}$: 

\begin{equation*}
\footnotesize
\label{master}
\begin{array}{r@{\hspace{-1.5mm}}l@{\hspace{-1mm}}r@{\hspace{-0.25mm}}l}
4t & \multicolumn{3}{l}{\frac{d}{dt}(E_{kl}) =}                          \\
\vspace{0.25mm}
  &   & (1-\alpha) & \left [ E_{k-2,l}(t)(k-2) + E_{k,l-2}(t)(l-2) + 2N_{l-2}(l-2)\delta_{k,1} \right ] + \\ 
\vspace{0.25mm}
  &  & \alpha     & \left [ E_{k-1,l}(t)(k-1) + E_{k,l-1}(t)(l-1) +  \enspace N_{l-1}(l-1)\delta_{k,2} \right ] - \\
\vspace{0.25mm}
  &  &            & E_{kl}(t)(k+l),                                    \\
\end{array}
\end{equation*}
where $\delta_{i,j}$ is the Kronecker delta.
Substituting $E_{kl}(t)=2 t e_{kl}$ and $N_k(t)=2 t p_k$ eliminates time from this equation and 
yields expressions for $e_{k\ge3,l} , e_{1,l\ge3}$ and $ e_{2,l\ge3}$.
%
To initialize the recurrences we similarly calculate
$e_{11}=0, e_{21}=e_{12}=2\alpha/7$, and $e_{22}=\alpha(e_{12}+p_1)/8$. 
In addition, because of the symmetry of the $E_{ij}$ terms,
and since the edges are undirected when $i=j$, we are over-counting so we divide $e_{ij}$ by 2. 
Conversely, when $0<|i-j|\le2$, we are under-counting and so we multiply $e_{ij}$ by 2.
Therefore, the $e_{kl}$'s (and hence, $\rho$) can be calculated~\cite{unpublished} for
a given value of $\alpha$. A plot of $\rho$ versus 
$\alpha$ can be seen in Fig.~\ref{figure:dd-assort}, 
%
where a good agreement is apparent with networks simulated from our model.  
Previous attempts to create a model that admitted varying $\rho$ values worked by rewiring the edges
of an existing network~\cite{sokolov:2004assort}.  
In contrast, our model grows
networks with a range of negative and positive $\rho$ values from first principles, giving insight into how assortativity may arise in 
networks.  We note that in our experiments $\rho$ approaches $0$ from the negative side for $\alpha<2/3$, 
but it does so very slowly and is negative for all networks we tried (up to $10^7$ nodes).
It can be shown that~\cite{unpublished} in the thermodynamic limit 
Newman's original formula for $\rho$~\cite{newman:02ami} yields $\rho=0$
when $\alpha < 2/3$.

\begin{figure}[t]
\center{
\includegraphics[width=0.80\columnwidth]{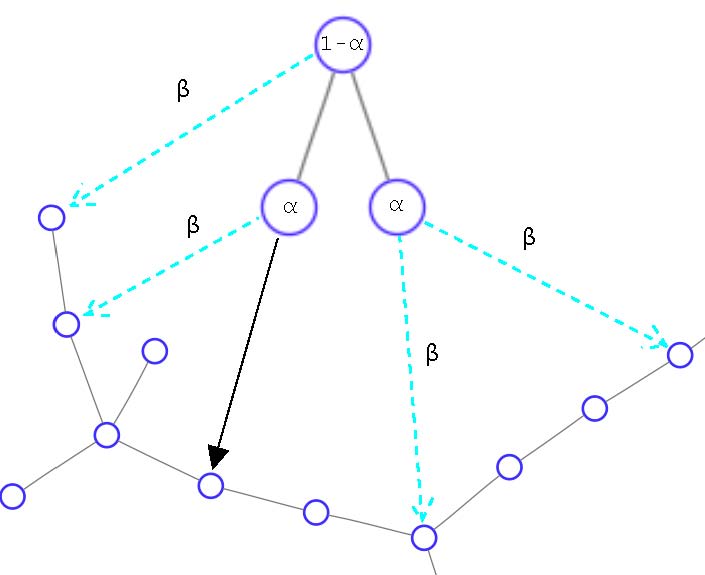}
\caption{
\label{figure:veebeta}  
(Color online) Illustration of the $\bigvee_\beta$ model. Once the graphlet attaches
to the network, based on $\alpha$, we introduce up to $l$ (here $l=4$) additional edges from the graphlet into random nodes of the network, each with probability $\beta$. }}
\end{figure}

The $\bigvee$-graphlet arrival model always produces trees and hence is not
expected to match empirical networks on some interesting properties (such as
clustering coefficient).  
Therefore, we examine a simple extension to the $\bigvee$-model 
which allows it to produce denser graphs,
without significantly affecting the model's degree distribution and
assortativity features.  The extended model,  illustrated in Fig.~\ref{figure:veebeta}, 
adds with probability $\beta$ at
each time step, {\bf {$l$}} edges (or dyads) from the arriving graphlet into the
existing network, with the attachment points being chosen uniformly at random.  In
addition to allowing denser graphs, this ``$\bigvee_\beta$-model" also reflects the
behavior in various real-world networks, where a newly arriving graphlet may
attach to the existing network at more than one point (\eg, new families
arriving in a city, etc.).
A theoretical analysis for the extended model is very complex. Instead, in the following model 
comparison we simulate networks for many values across the possible parameter space
$(\alpha,\beta, l)$.

Existing  network literature compares networks or network models by 
studying one or two particular properties, and most commonly the degree 
distribution. In this letter, we introduce a fifteen-dimensional attribute vector 
of seven well-known network properties, which should enable a general and comprehensive 
comparison between any set of networks.  
These properties are: 
the number of nodes, the number of edges, the geodesic
distribution, the betweenness coefficient distribution, the clustering
coefficient distribution, the assortativity, and the degree distribution of the
network.  For the four distributions, we use the mean, standard deviation, and
skewness as proxy attributes, for a total of $15$ attributes.  Networks 
are mapped to points in a $15$-dimensional space defined by these attributes, normalizing each value by subtracting the attribute mean and dividing by the attribute's standard deviation.
%

\begin{figure}[t]
\includegraphics[width=0.96\columnwidth]{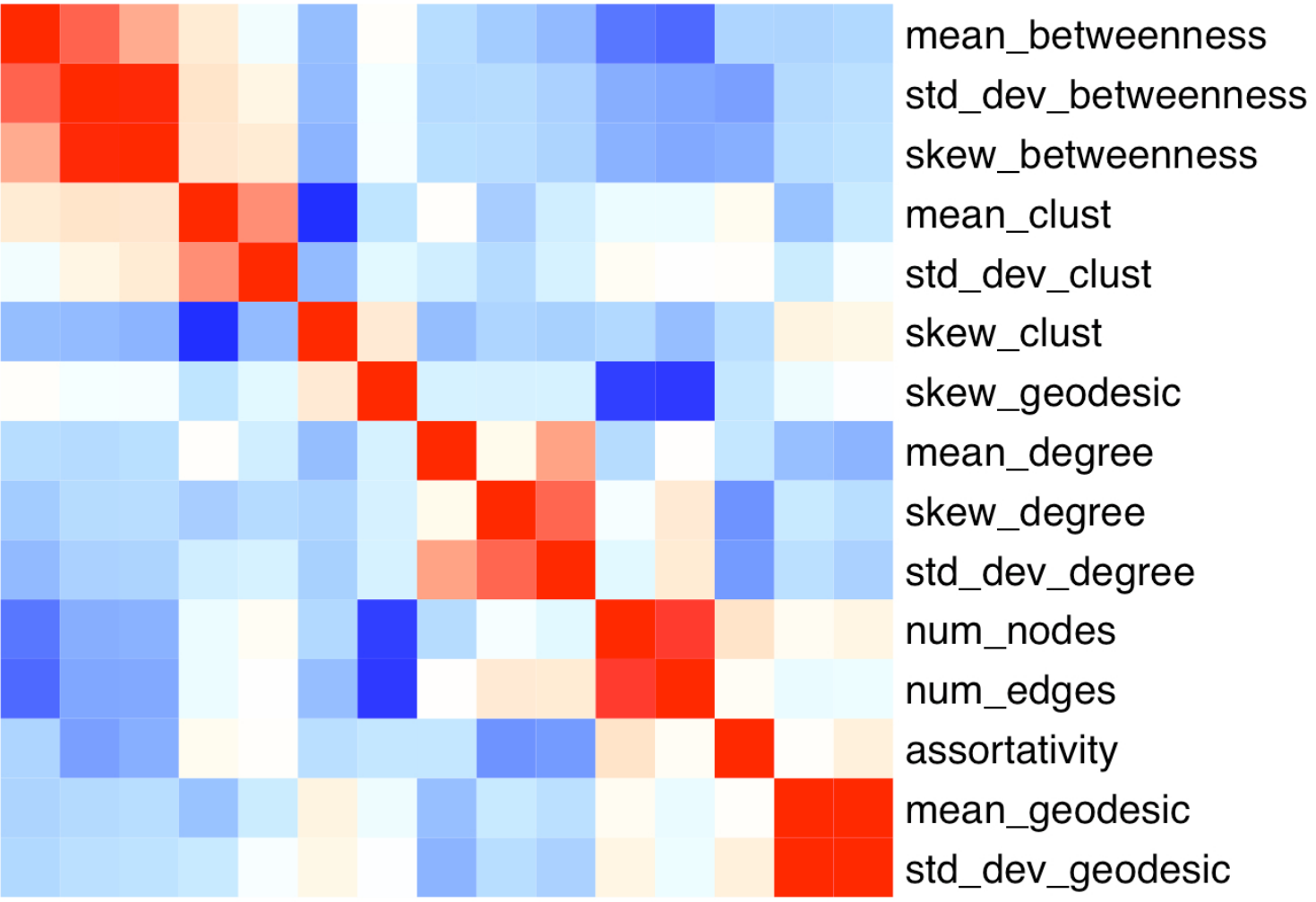}
\caption{
\label{figure:heatmap}  (Color online) Symmetric heatmap of attribute correlations among
networks.  Red (blue) indicates perfect correlation (anti-correlation). 
White is the intermediate case of no correlation. The
small amount of clustering along the diagonal attests to the relative
independence of the attributes. }
\end{figure}

Our collection
of real-world networks consists of $113$ diverse networks from biological, social
and technical domains.  
It includes software call
graphs~\cite{chris:apache}, a social network of software
developers~\cite{chris:email}, political social networks~\cite{newman:polblogs,
newman:polbooks}, 
$3$ gene networks~\cite{alon:coli, yeastract, lee:chip-chip}, 
$3$ protein-protein
interaction networks~\cite{biogrid}, cellular networks for several
organisms~\cite{barabasi:cellular}, and several others downloaded from a web
repository of networks~\cite{newman:data-page, newman:adjnoun,
newman:neural-power, newman:collab, newman:dolphins, newman:football,
newman:karate, newman:lesmiz}.
The degree of overlap, or dependence, between the attributes when
characterizing networks can be assessed by the symmetric heatmap in
Fig.~\ref{figure:heatmap}, showing the pairwise correlations (Pearson) of the
network attributes over a representative sample of real-world networks (one
from each data set described above).  The rows and columns of the heatmap are
ordered so that, within the limitations of the hierarchical clustering used, 
the attributes most correlated with each other are placed closest.
The map allows us to identify clusters of ``similar'' network attributes by
looking for blocks of squares along the diagonal of the figure.
Since there is only a small amount of clustering along the diagonal, it follows
that most network attributes we have chosen are relatively independent, and
thus, provide information to our analysis.

In the following analysis we have eliminated 4 of the 15 network attributes and retained 11.
One reason, is that some attributes, like number of nodes and edges were tightly correlated as indicated in the heatmap.
So we only kept one of them, the number of edges.
Another reason is that since the $l$'th moment of a power-law distribution, $p(k)\sim k^{-\gamma}$, is only defined for  $l  < (\gamma-1)$,  we have omitted the skew of the degree distribution, as a precaution. 
For the same reason, the variance and skew of the betweenness distribution have been left out, even though the exact nature of the betweenness distribution does not seem to be known. The distribution of the  clustering coefficient and geodesic are defined for the models investigated in this paper~\cite{dorgov2}  and  have hence been retained.

Next, we compare a collection of $\bigvee_\beta$-arrival growth networks to the above 
collection and to a baseline collection of networks from the well-known BA model~\cite{barab:albert}.
We chose BA as a baseline because, like BA, our graphlet-arrival model uses the mechanism of preferential attachment, only instead of nodes we have graphlets arriving.  We sample a large swath of the parameter space for the $\bigvee_\beta$-arrival model, iterating across several possible values for each parameter and creating networks that cover the size range of real-world networks.  To this end, we use network sizes ranging from $500$ to $5250$ nodes at $250$ node intervals, $\alpha$ values in the range $0\le \alpha \le 1$ at
intervals of $0.1$, $\beta$ values in the range $0\le \beta \le 1$ at intervals of $0.1$, and $l$ values in the range $1$ to $5$.  For each possible combination of values of these four parameters, we create five networks, giving us a total of $60,500$ networks.  For the BA model, we generate $500$ sample networks by varying the number of nodes in the same range as our model (with identical increments), varying the number of edges added at each attachment from $1$ to $5$, and creating $5$ sample networks for each possible combination of these two parameters.  

\begin{figure*}
\includegraphics{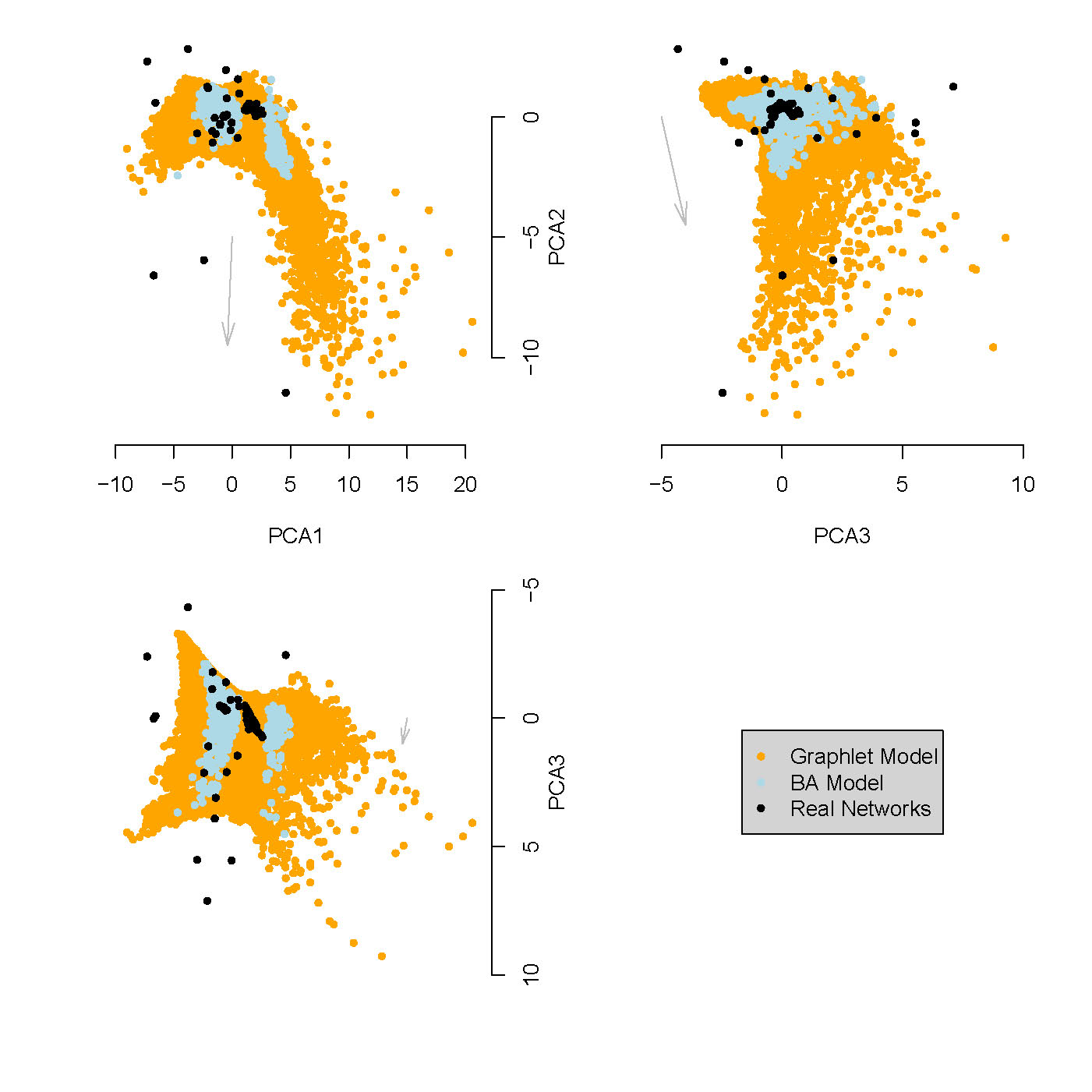}
\caption{ \label{figure:pca} (Color online) A projection of our model nets (orange points), the BA model nets (light blue), and real-world nets (black) onto the first three principal components of the eleven dimensional PCA space of our real-world data set
(we omitted the number of nodes, skew of the degree distribution, and variance and skew of the betweenness distribution from the original 15 attributes, as described in the text).
Here, the PCA1 axis is primarily composed of (in terms of their coefficient's magnitude) a combination of the 
number of edges, mean and skew of geodesic,
mean and st. dev. of clustering, and mean and st. dev. of degree.
PCA2 is mainly a combination of the st. dev and mean of geodesic, and assortativity.
PCA3 is mainly a combination of the mean of betweenness, mean of clustering, number of edges, and st. dev of degree.
As an example of a spread along an original parameter, the grey arrow is parallel to and shows the direction and magnitude of assortativity when projected onto this space.
}
\end{figure*}

To objectively assess the extent to which our model networks cover the range of attributes simultaneously, we visualize the attribute space using an established statistical 
dimension-reduction technique, {\em Principal
Components Analysis} (PCA), which guarantees maximal retention of the variance when
projecting data into a lower dimension~\cite{jolliffe:pca}.  PCA finds the
projection of an $n$-dimensional data set onto a space of the same dimension,
where the new axes, or {\em principal components}, are orthogonal and linear
combinations of the original dimensional variables, such that the first $d$
axes, $d \le n$, retain the maximal variance of the original data set possible
with that many dimensions.   
Fig.~\ref{figure:pca} shows the projections of the sets of $\bigvee_\beta$ model, BA model, and real-world networks onto the first three principal components (out of 11) of the real-world data set found by the PCA algorithm.
These principal components retain $71 \%$ of the original data variance and demonstrate the larger coverage potential of the extended graphlet arrival model.
We note that these results are fairly stable with respect to the number of variables used in the PCA analysis:
using between $2-4$ fewer (or more) than the 11 variables does 
not qualitatively change the results~\cite{unpublished}.
While PCA has been used before to cluster networks~\cite{dacosta},
our methodology here is novel in that it offers a general and explicit way to compare growth models relative to each other, with respect to the fraction of PCA space they cover. 
Additionally, it allows for models to be compared more finely, along individual or combinations of original variables,
by projecting those variable vectors in the same PCA space, e.g. assortativity in Fig.~\ref{figure:pca},
and then observing the spread differentials between the model networks along those vectors. 
 

In conclusion, graphlet arrival models are a positive step toward more realistic network models
which, as we show, better approximate empirical networks in biology, software, and social science, both
in the modeling step (graphlet versus node arrival) and in the results
(matching more complex measures of networks, like assortativity).  A broad
degree distribution and wide variation of assortativity are features of the
$\bigvee$-arrival model which are not present in preferential attachment models
that grow via individual nodes, and/or edges.  In particular, we believe that
the attachment asymmetry of the $\bigvee$-graphlet is largely responsible for
these features and that they would not be apparent in a graphlet model of fully
connected graphlets (e.g edge, triangle, or square).  
Therefore, we expect more complete graphlet arrival models (whose theoretical analysis 
would also be more complex), considering a larger set of possible graphlets 
to yield even better models of empirical networks 
(we also note that the addition of simpler graphlets should expand the range of possible 
$\gamma$'s to below $3$, where the exponents of most real-world networks with power-law degree-distributions reside). Finally, we anticipate  that 
the technique of comprehensive comparison of  networks across a suite of network 
properties introduced in this letter, would find wider use in the network literature.

This work was funded in part by the National Science Foundation under Grant No. IIS-0613949.
We thank the anonymous referees for their invaluable comments which improved this paper.

\end{document}